# From Stars to Patients:  Lessons from Space Science and Astrophysics for Health Care Informatics

S. G. Djorgovski, A. A. Mahabal
California Institute of Technology
Pasadena, CA 91125, USA
[george,aam]@astro.caltech.edu

B. Chaudhry
TupleHealth
Washington, DC 20008, USA
basitchaudhry@tuplehealth.com

D. J. Crichton
Jet Propulsion Laboratory
Pasadena, CA 91109, USA
daniel.j.crichton@jpl.nasa.gov

*Abstract*—Big Data are revolutionizing nearly every aspect of the modern society.  One area where this can have a profound positive societal impact is the field of Health Care Informatics (HCI), which faces many challenges. The key idea behind this study is: can we use some of the experience and technical and methodological solutions from the fields that have successfully adapted to the Big Data era, namely astronomy and space science, to help accelerate the progress of HCI? We illustrate this with examples from the Virtual Observatory framework, and the NCI EDRN project.  An effective sharing and reuse of tools, methods, and experiences from different fields can save a lot of effort, time, and expense. HCI can thus benefit from the proven solutions to big data challenges from other domains.

## I. The Goals of This Study

Data intensive systems are now playing a key role in enabling science, from improving the efficiency of automating capture, management and scientific analysis of data to ensuring the digital preservation of the scientific record of research.  The advent of Big Data in scientific research as well as many other fields is rapidly making traditional approaches to managing and analyzing data obsolete.  The volume, distribution, heterogeneity and complexity of the data require new paradigms for its generation, management, distribution and analysis.  It is now broadly recognized that systematizing the data generation and capture across the entire research lifecycle is critical in enhancing scientific collaboration, enabling analyses using advanced computational methods, and supporting reproducibility of scientific results [1].

Similar challenges and opportunities are present in other communities of practice, such as the Health Care Informatics (HCI).  This is a heterogeneous arena that spans a variety of goals and constituencies, ranging from a basic biomedical research to a practical/clinical HC delivery, HC management, policy, and commercial interests.  This suggests that a variety of solutions will be needed, and there are many challenges [2].

## II. A Brief Overview of the Health Care Informatics

We explored the existing HCI literature to identify some specific areas or challenges in HCI, where a methodology transfer may be useful and relatively quick to accomplish. Likewise, we explored the currently publicly available HCI data sets to identify the areas where improved practices are needed in order to make an effective use of the modern data science tools.

The diversity of HCI needs, methods, etc., is reflected in the literature. The data are not just diverse, but of diverse types, including patient data, disease related data, data of effectiveness of treatments, HC management data, customer relationships, etc., and traditional methods are insufficient [3].  Another facet of diversity is the different levels at which data are collected: molecular, tissue, patient, and population [4]. Translational BioInformatics (TBI) exploits data from the different levels, is an important direction for HCI [5].

The biggest current hurdle in our ability to use data mining on many datasets is the lack of metadata, the necessary requirements of anonymizing patient data before it can be merged with other relevant datasets, and this in turn resulting in a lack of large uniform and interoperable datasets. This further hampers reproducibility of results on both levels: from patient to data collection, and from data to results.  There is an increasing attention to interoperability and standards [6]. While the lack of interoperability that arises from patient privacy concerns and legal requirements may hinder meaningful data mining, there is also sometimes information blocking, when persons or entities interfere with the exchange or use of electronic health information, due to economic interests or management issues [7].

In general, well documented, complete, homogeneous data sets are hard to find, and they are mainly focused on particular projects or narrow issues. Some of the noteworthy efforts include the Clinical Research Information System (CRIS) that contains nearly 90% of medical records from NIH's Clinical Center [8], and the development of NIH's

Biomedical Translational Research Information System [9]. The National Library of Medicine (NLM) recently announced its strategic vision to: (a) catalyze data sharing and reproducibility, (b) be the programmatic epicenter of Data Science at NIH, and (c) make accessible nation's medical legacy [10]. A number of relevant articles can be found in [11], especially [12]. Some of the useful data depositories and collections include [13,14]. The Observational Health Data Sciences and Informatics (OHDSI; http://www.ohdsi.org) is especially noteworthy, as it also includes a number of useful data tools.

III. VIRTUAL OBSERVATORY AS AN EXAMPLE OF A DOMAIN-SPECIFIC CYBERINFRASTRUCTURE

Virtual Observatory (VO) in particular provides a good model of an organizing framework for Big Data science. Parallels between VO and the data challenges in genomics have been discussed in [15].

VO was the astronomy's community response to the challenges and opportunities on the exponential data avalanche. It was envisioned as a complete, distributed (Web-based) research environment for astronomy with large and complex data sets. It was meant to federate geographically distributed data and compute assets, and the necessary expertise for their use. Already by the 1980's, most astronomical data were collected in a digital form (and the older, photographic data were soon digitized). By mid-1990's, astronomy entered the Terascale regime, due to the large digital sky surveys, and it is in the Petascale regime now, with the exponential growth continuing. The concept of the VO framework was formulated around the year 2000 [16,17,18], and it was quickly embraced by the community world-wide [19].

VO facilitates discoveries in several ways. First, it provides a unified way to access data from a broad variety of sources. The sheer size of the data sets, often containing $\sim 10^8 - 10^9$ sources (stars, galaxies, quasars, etc.) with $\sim 10^2 - 10^3$ measured parameters each enables statistical studies with a great precision, and the discovery of very rare types of objects. Data fusion (e.g., between different wavelength regimes) often reveals knowledge that is present in the data, but not recognizable in any of the data sets separately. The high information content of the data enables an effective data mining, with the same data being used for a variety of different projects. Open data access allows anyone with an internet connection to do a first rate science, thus engaging a broader pool of talent and fostering the democratization of science.

An essential element was a growing web of collaborations between the big-data, computationally savvy astronomers, and applied computer scientists. Both groups need to learn each other's language, to understand each other's priorities and expectations, and different styles of research. The same phenomenon plays out on an institutional level as well, with the different styles, goals, and value systems.

There are several lessons learned. First, it is important to understand the capabilities and the willingness of the constituent user community to change their mode of research, make strong efforts to educate the community, make their tools and services easy to learn and use, and nurture a steady stream of actual scientific results. Riding the exponential wave of Moore's requires an institutional agility, adopting to a rapidly changing technology, and the changing scientific landscape. There is also an inherent tension between an exploratory and innovative spirit and creative people who embody it, and a strong management and a well defined organizational structure and administrative requirements.

Today, VO is the global data grid of astronomy, and is regarded as one of the success stories of the virtual scientific organizations and cyber-infrastructure. This was due to a number of factors:

- All data are collected in a digital form.
- Computing- and data-savvy community.
- Some standard formats were already in place.
- Large data collections are in funded, agency mandated, publicly accessible archives.
- There is an established culture of data sharing and reuse.
- VO was a community initiative driven by the needs of an exponential data growth.
- Federal agency support and funding were available.
- Data have no commercial value or privacy issues.

Essentially all of these conditions are currently limited or absent in the HCI arena, and therein lies the challenge.

IV. A SUCCESSFUL METHODOLOGY TRANSFER EXAMPLE

We illustrate this approach with a successful example of methodology transfer from space science to medicine: Using a state-of-the-art informatics infrastructure developed at JPL and leveraging successful efforts to build similar open source systems in space science open-source Object Oriented Data Technology (OODT; http://oodt.apache.org) to design and an effective national integrated Knowledge System for the National Cancer Institute's Early Detection Research Network (EDRN; http://www.cancer.gov/edrn).

EDRN is an international collaboration comprising dozens of institutions focused on cancer biomarker research, has faced many big data challenges as a growing, distributed, scientific network. These challenges led to the development of the EDRN Knowledge System that integrates heterogeneous, distributed data and provides researchers the ability to capture, access, process and share data and knowledge generated by biomarker research efforts [20]. Using a state-of-the-art informatics infrastructure developed by NASA's Jet Propulsion Laboratory (JPL), an operating division of the California Institute of Technology, and leveraging successful efforts to build similar open source systems in space science, the EDRN was able to deploy an effective, integrated Knowledge System on national scale.

Much of the capabilities in EDRN build on the success of JPL's experience in supporting data intensive science for NASA. For example, we heavily leveraged many of the programmatic, technical, and scientific approaches employed by the Planetary Data System (PDS), a distributed set of data repositories archiving data from all U.S. robotic solar system missions and instruments formed after a National Research Council recommendation to address long-term digital data preservation and research challenges in space science. The leveraging between EDRN and JPL has allowed for data science capabilities developed for the EDRN to be re-infused back into NASA missions and projects as well.

The leveraging has directly focused on: structuring programs in large-scale data management and analysis; architecting national and international data intensive systems; ensuring software and data architectures, particularly for data intensive science, can evolve as science changes; developing scientific data standards for annotating scientific data sets; enabling peer review of scientific data to improve overall data quality; integrating data across the scientific data discovery process; unifying scientific data and publications; developing open source technologies that can be applied across domains; and working between government and academia to have a long-term model for both data analysis and preservation.

The software framework used to build these systems is built on top of the Apache OODT open source software stack (http://oodt.apache.org), originally developed at JPL to address construction of data intensive systems, provides the building blocks for constructing these systems. Beyond planetary and cancer research, it is being used to develop data intensive systems for climate and Earth Science as well as observational systems for defense agencies. Many of the challenges across these domains are similar, stressing the need for both integrating and scaling data processing, management and analysis.

V. CONCLUDING REMARKS

We conclude that data methodology transfer from astronomy and space science into HCI is clearly possible, at least within a given practical subdomain. However, the greater complexity and sociological issues will likely delay progress. Specifically, some of the outstanding challenges include:

- The HC community does not yet have a well-developed data culture and technical skills for data management and analytics; acquiring them will take some time and resources.
- The data and metadata must be understood, relevant, repeatable, with standard formats, properly curated, and with standard interoperability protocols.
- Diverse stakeholders (researchers, HC providers, commercial entities, government agencies) may have mutually incompatible interests, yet need to find common goals and work together.
- Adequate privacy protection mechanisms are needed that would not be stifling innovation and further progress.

We recommend the following path forward for this field:

- Promote further multidisciplinary collaborations between communities of practice that have developed large scale open data environments (e.g., astronomy, space science, physics) and HCI.
- Organize knowledge transfer activities through which best practices and lessons learned could be shared between research communities to help accelerate progress in HCI.
- Work with funding agencies to facilitate multidisciplinary collaboration around large scale HCI data analysis and infrastructure development.
- Facilitate development of training programs in HCI that draw on expertise from other domains.

ACKNOWLEDGMENT

This work was supported in part by a grant from the Electronic Data Methods Forum (EDMF; http://www.edm-forum.org), by the Center for Data-Driven Discovery at Caltech (CD$^3$; http://cd3.caltech.edu), and by the Center for Data Science and technology at JPL (http://datascience.jpl.nasa.gov).